\documentclass[%
reprint,prl,
superscriptaddress,
showpacs,
amsmath,amssymb,
aps,
]{revtex4-1}

\usepackage{dcolumn}
\usepackage{bm}
\pdfpageattr{/Group << /S /Transparency /I true /CS /DeviceRGB>>} 
\usepackage{amsmath}
\usepackage{amssymb}
\usepackage{amsfonts} 
\usepackage{wasysym}  
\usepackage{graphics}  
\usepackage{psfrag} 
\usepackage{graphicx} 
\usepackage{epsfig}    
\usepackage{rotating}
\usepackage[latin1]{inputenc}
\usepackage{color}
\usepackage{rotating}
\usepackage{csquotes}
\usepackage{multirow}
\usepackage{nicefrac}
\usepackage[nooneline, tight,raggedright,FIGTOPCAP]{subfigure}
\usepackage[linkcolor=red,citecolor=blue,urlcolor=blue,colorlinks=true]{hyperref}
\usepackage{color}
\usepackage{epsfig}
\usepackage{comment}
\definecolor{darkgreen}{rgb}{0,0.5,0} 
\definecolor{violet}{rgb}{0.5,0,0.5}
\definecolor{orange}{rgb}{0.2,0.5,0.5}

\newcommand{\e}{\mathrm{e}}
\newcommand{\imag}{\mathrm{i}}

\newcommand{\summe}[3]{\sum\limits_{#1 = #2 }^{#3}\;}
\newcommand{\f}{f}

\newcommand{\bequ}{\begin{equation}}
\newcommand{\eequ}{\end{equation}}
\newcommand{\bequa}{\begin{eqnarray}}
\newcommand{\eequa}{\end{eqnarray}}
\newcommand{\bse}{\begin{subequations}}
\newcommand{\ese}{\end{subequations}}

\newcommand{\Pft}{\hat{P}_k}
\newcommand{\Pftnull}{\hat{P}_{k}}
\newcommand{\rhohom}{\bar{\rho}}

\renewcommand{\vec}[1]{\mathbf{#1}}
\newcommand{\ie}{i.\,e.\ }
\newcommand{\eg}{\textit{e.g. }}

\newcommand{\Real}{\mathrm{Re}}
\newcommand{\Imag}{\mathrm{Im}}

\begin{document}


\title{Active Curved Polymers form Vortex Patterns on Membranes}
\author{Jonas Denk}
\author{Lorenz Huber}
\author{Emanuel Reithmann}
\author{Erwin Frey}
\email{frey@lmu.de}
\affiliation{
Arnold Sommerfeld Center for Theoretical Physics (ASC) and Center for NanoScience (CeNS), Department of Physics, Ludwig-Maximilians-Universit\"at M\"unchen, Theresienstrasse 37, D-80333 M\"unchen, Germany
}

\begin{abstract}
Recent \textit{in vitro} experiments with FtsZ polymers show self-organization into different dynamic patterns, including structures reminiscent of the bacterial Z-ring. We model FtsZ polymers as active particles moving along chiral, circular paths by Brownian dynamics simulations and a Boltzmann approach. Our two conceptually different methods point to a generic phase behavior. At intermediate particle densities, we find self-organization into vortex structures including closed rings. Moreover, we show that the dynamics at the onset of pattern formation is described by a generalized complex Ginzburg-Landau equation.
\end{abstract}

\maketitle

Intracellular structuring is often facilitated by the active dynamics of cytoskeletal constituents.
The origin of these driven dynamics and their impact on pattern formation has been extensively studied using artificial motility assays of cytoskeletal filaments~\cite{schaller_frozen_2011,Schaller2010,Sumino2012,suzuki_polar_2015}. 
Another intriguing example of self-organization due to driven filaments was reported recently by Loose and Mitchison~\cite{loose_bacterial_2013} \textit{In vitro}, the bacterial protein FtsZ forms membrane-bound, intrinsically curved polymers. These seem to exhibit treadmilling dynamics (consuming GTP) and, as a result, move clockwise on the membrane.  Depending on the protein density, polymers cluster into dynamic structures such as rotating rings or jammed bundles, despite the absence of attractive interactions~\cite{meier_form_2014}. These ring structures are of particular interest, since \textit{in vivo}, FtsZ builds the contractile Z-ring which drives cell division in a yet unknown way~\cite{Ingerson-Mahar2012,erickson_ftsz_2010,szwedziak_architecture_2015}. But also in the \textit{in vitro} experiments, the pattern forming mechanism remain unclear even on a qualitative level. 

\begin{figure}[t]
\centering
\includegraphics[width=0.9\columnwidth]{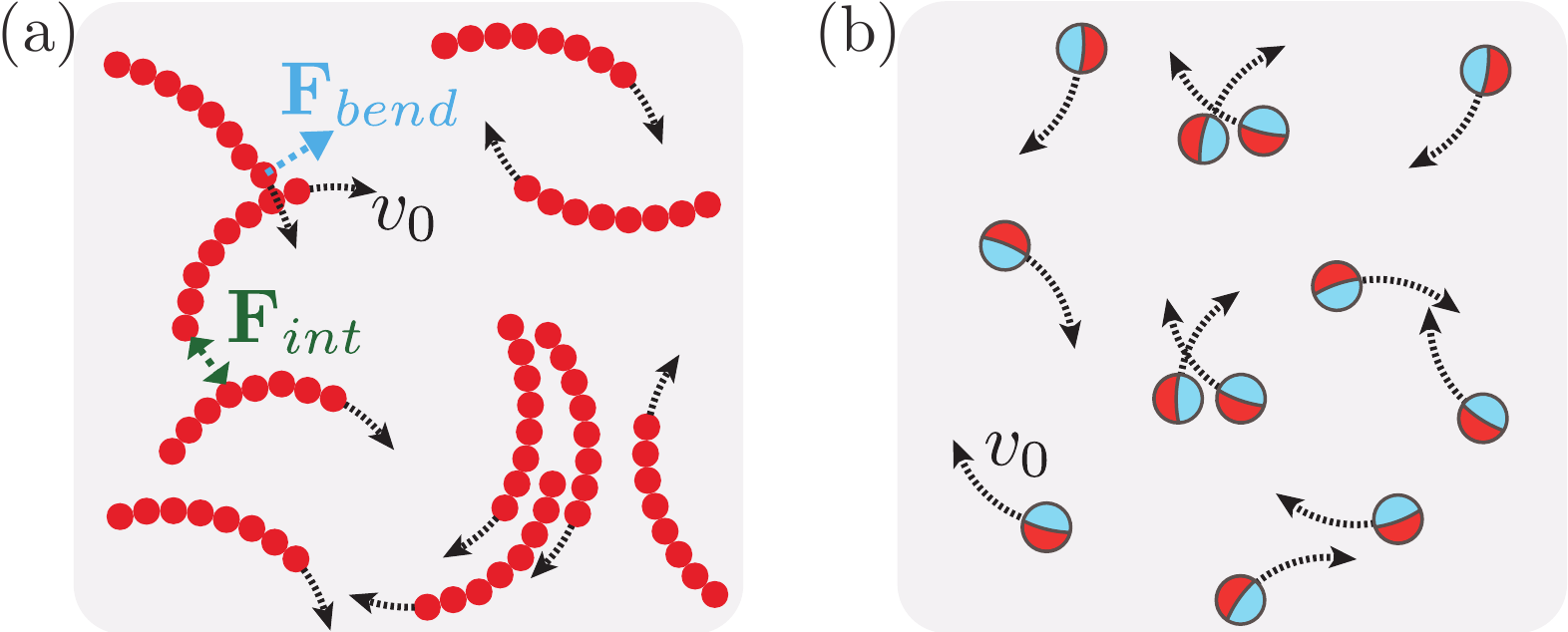}
\caption{Systems of active particles, which are driven on chiral, circular tracks with speed $v_0$: \textbf{(a)} \emph{Microscopic view}: extended, elastic polymers with intrinsic curvature, where noise and steric interaction trigger bending of filaments. \textbf{(b)} \emph{Mesoscopic view}: point-like particles that undergo diffusion as well as binary collisions.}
\label{Fig::illustration}
\end{figure}

Motivated by these experimental findings, we have studied pattern formation in a class of active systems, where particles move on circular tracks and interact only via steric repulsion. To assess the dynamics of this class, we consider two conceptually different models: First, we emulate active particles as elastic polymers with fixed intrinsic curvature that move with a constant tangential velocity [Fig.~\ref{Fig::illustration}(a)] and perform Brownian dynamics simulations.  Second, we employ a kinetic Boltzmann approach, where point-like particles move on circular paths and undergo diffusion and binary collisions (with polar symmetry) according to a simplified collision rule [Fig.~\ref{Fig::illustration}(b)]. As a result, we identify different phases of collective behavior as a function of density and noise level. With both approaches, we find flocking into vortex patterns in the regime of intermediate density and noise strength. Our simulations for extended particles predict the formation of closed ring structures reminiscent of those found in Ref.~\cite{loose_bacterial_2013}, even in the absence of any attractive interactions. In the mesoscopic limit, our analysis yields that, close to the onset of vortex formation, the dynamics at onset of ordering is characterized by a novel generalization of the complex Ginzburg-Landau equation.

In our Brownian dynamics simulations we consider a system of $M$ curved polymers of the same chirality embedded in a two-dimensional membrane of area $A$ with periodic boundary conditions. Each polymer is described as an inextensible worm-like chain~\cite{kratky_rontgenuntersuchung_1949, saito} of length $L$, persistence length $\ell_p$, and intrinsic curvature $\kappa_0$. For a given polymer conformation $\vec{r}(s)$, parameterized in terms of arc length $s$, the overall bending energy is given by $E_\text{bend} {=} \frac12 \ell_p k_B T\int_0^L\mathrm{d}s\left[\kappa(s){-}\kappa_0\right]^2$, where $\kappa(s) {=}|\partial_s^2 \vec{r}(s)|$ denotes the local curvature. Excluded volume interaction is implemented by a repulsive truncated Lennard-Jones potential (for details see the Supplemental Material~\cite{SUP}). To assure motion of the filament contour on a circular track (apart from noise), polymers are propelled with a tangential velocity $\vec{v}_0(s){=}v_0\partial_s\vec{r}(s)$. This accounts for the effective motion of treadmilling in a simplified way~\cite{SUP}. Note that for this choice, the area explored by a circling polymer is minimal. In the free draining limit, the dynamics of the polymer system is then determined by a set of coupled Langevin equations for the contours $\vec{r}^{(m)}(t,s)$ of each polymer $m{=}1,2...,M$: 
$\zeta \bigl( \partial_t\vec{r}^{(m)}{-}\vec{v}^{(m)}_0 \bigr){=}{-}{\delta E [\{ \vec{r}^{(n)} \}]}/{\delta\vec{r}^{(m)}}{+}\boldsymbol{\eta}^{(m)}$, balancing viscous friction with elastic and repulsive forces generated by the total energy $E$ and Langevin noise $\boldsymbol{\eta}$ with zero mean and $\langle\boldsymbol{\eta}(t,s)\cdot\boldsymbol{\eta}(t',s')\rangle{=}4k_BT\zeta\delta(t-t')\delta(s-s')$. 
To numerically solve the polymer dynamics we employ a bead-spring representation of the polymers~\cite{chirico_kinetics_1994, wada_stretching_2007}. For most simulations, we adapted length scales close to those observed in Refs.~\cite{loose_bacterial_2013, erickson_ftsz_2010}: $\kappa_0^{-1} {=} 0.5\,\mathrm{\mu m}$, $L {=} 0.9\,\mu m$, $\ell_p {=} 10\,\mu m$. The relevant dimensionless parameters that characterize the system are the reduced noise ${ \sigma}$ and density $\rho$. Here, ${\sigma}{:=}k_BT\ell_p/(\zeta v_0 L^2)$ relates thermal forces at length scale $\ell_p$ with friction forces, and $\rho {:=}(R_0/b)^2$  denotes the squared ratio of the radius of curvature $R_0{=}\kappa_0^{-1}$ to the mean polymer distance $b{=}\sqrt{A/M}$.

For dilute systems, $\rho {\ll} 1$, our simulations show that each polymer is propelled on a circular path and collisions between polymers are infrequent; see Fig.~\ref{Fig::resultsBD}(a) and Movie 1 in the Supplemental Material~\cite{SUP}. 
The positions of the polymers' centers of curvature $\vec{r}^{(m)}_{cc}$ are uncorrelated as in a gas, and we refer to this state as a \emph{disordered state}. 
On increasing ${ \rho}$, we observe that a significant fraction of filaments begin to collide and collect into localized vortex structures (\textit{vortex state}). 
These ring-like structures are highly dynamic. They assemble and persist for several rotations, during which their centers of mass remain relatively static; see Fig.~\ref{Fig::resultsBD}(b) and Movie 2~\cite{SUP}. Despite our simplified kinetic assumption, the overall phenomenology resembles the FtsZ patterns observed by Loose and Mitchison~\cite{loose_bacterial_2013}, including vortex assembly, disassembly and localization. 
In the dense regime, ${ \rho} {\gtrsim} 1$, where each polymer is likely to collide, these vortices are unstable. Instead, the polymers cluster and form jammed `trains' that travel through the system in an irregular fashion; see Fig.~\ref{Fig::resultsBD}(c) and Movie 3~\cite{SUP}. 

In order to quantitatively distinguish between the various observed patterns and organize them into a `phase diagram' we consider the pair correlation function $g(d_{cc})$~\cite{dhont_introduction_1996, lu_molecular_2008} of distances $d_{cc}{=}|\vec{r}^{(m)}_{cc}{-}\vec{r}^{(n)}_{cc}|$ between the centers of curvature [Fig.~\ref{Fig::resultsBD}(d)]. We regard a system as \emph{disordered} if $g(d_{cc})$ exhibits a minimum at a distance $d_{cc}^\text{min}$ equal to the diameter of a free circular path, $d_{cc}^\text{min}{\approx} 2 R_0$. This is distinct from \emph{vortex states}, where $d_{cc}^\text{min}$, defining an effective vortex diameter, is larger than $2R_0$. Finally, for \emph{train states}, $g(d_{cc})$ does not exhibit a local minimum, indicating the absence of an isolated vortex structure; for more details see the Supplemental Material~\cite{SUP}. 
\begin{figure}[t]
\centering
\includegraphics[width=\columnwidth]{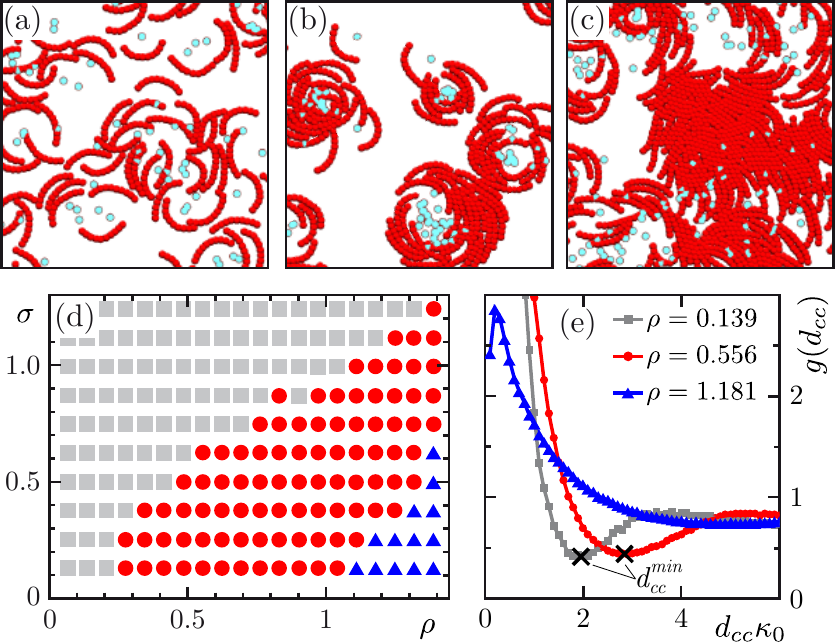}
\caption{System snapshots are provided to depict \textbf{(a)} disorder ($\rho{=}0.556$, $\sigma{=}0.987$), \textbf{(b)} vortices ($\rho{=}0.556$, $\sigma{=}0.247$) and \textbf{(c)} trains ($\rho{=}1.389$, $\sigma{=}0.247$). Curvature centers $\vec{r}^{(m)}_{cc}$ are depicted by light blue dots. \textbf{(d)} Phase portrait for varying density $\rho$ and noise $\sigma$: \textit{disorder} states (gray rectangles), \textit{vortex} states (red circles), \textit{train} states (blue triangles). \textbf{(e)} Pair correlation function $g(d_{cc})$ for the three different states with $\sigma{=}0.247$ and $\rho$ indicated in the graph.}
\label{Fig::resultsBD}
\end{figure}
The ensuing `phase diagram' is shown in Fig.~\ref{Fig::resultsBD}(d). As in other active systems~\cite{Vicsek1995, Riedel2005, Chate2008, Bertin2008, ihle_kinetic_2011, Ginelli2010, peruani_collective_2012, Marchetti}, pattern formation is favored by increasing density and decreasing noise strength. 
Jammed states prevail only when density is high and noise level low. 
Note also that the structure of the phase diagram depends on the ratio of filament length $L$ to radius of curvature $R_0$. 
Polymers with an arc angle close to $\kappa_0 L {=} 2\pi$ (closed circles) retain a single-circle structure and do not form any collective structures upon increasing $\rho$ [Movie 4~\cite{SUP}]. Conversely, reducing $\kappa_0 L$ suppresses the formation of closed ring structures, due to inefficient alignment of short polymers. Instead, these polymers cluster into flocks which move on approximately circular paths [Movie 5~\cite{SUP}]. Hence, we conclude that the range of arc angles  of FtsZ polymers, $\kappa_0 L {\approx} 0.6 \pi$, observed \textit{in vitro}~\cite{loose_bacterial_2013}, facilitates the formation of closed polymer rings particularly well [Fig.~\ref{Fig::resultsBD}(b)]. In summary, closed polymer rings require explicit curvature and filament lengths larger than a certain threshold value. For other interactions than local, steric repulsion ring structures may also emerge~\cite{sumino_large-scale_2012,schaller_frozen_2011,yang_self-organized_2014}; straight, rotating rods may form vortex arrays but not closed rings~\cite{kaiser_vortex_2013}. 

We complement the Brownian dynamics simulations of active particles that are propelled on circular tracks by considering the mesoscopic limit of vanishing particle extension. To this end, we have employed a kinetic Boltzmann approach~\cite{BertinBoltzmannGLApproach, BertinLong, Bertin2008, Peshkov, FloSNAKE, FloParticleConservation, FloCriticalAssessment} to determine the collective behavior and the corresponding phase transitions in this limit, irrespective of the microscopic details of the constituent particles. In detail, we simplified the active system to one consisting of spherical particles (of diameter $d$) moving clockwise with constant speed $v_0$ on circular orbits of radius $R_0$. This accounts for both self-propulsion and spontaneous curvature but neglects the finite extension of the polymers as compared to our Brownian dynamics simulations. 

We further assume that a particle's orientation is altered by `self-diffusion' as well as by local binary collisions. 
In self-diffusion, a particle's instantaneous orientation $\theta$ changes at rate $\lambda$ into $\theta{+}\eta$, where we assume $\eta$ to be Gaussian-distributed with standard deviation $\sigma$. 
As in other particle-based active systems \cite{BertinLong,Peshkovmetricfree,FloSNAKE}, binary collisions are modeled by a polar alignment rule where the orientations of the collision partners align along their average angle plus a Gaussian-distributed fluctuation; for simplicity, we take the same width $\sigma$ as for self-diffusion.

The kinetic Boltzmann equation \cite{BertinBoltzmannGLApproach, BertinLong, Bertin2008, Peshkov, FloSNAKE, FloParticleConservation, FloCriticalAssessment} for the one-particle distribution function $f(\vec{r},\theta,t)$ then reads
\begin{align}
\label{Eq:Boltzmannactivematter}
&\partial_tf{+}v_0\big[\vec{e}_\theta{\cdot}\partial_{\vec{r}}{+}\kappa_0 \partial_\theta\big] f =\mathcal{I}_\textit{d}[f]{+}\mathcal{I}_\textit{c}[f,f]\,.
\end{align}
It describes the dynamics of the density of particles in phase-space element $\mathrm{d}\vec{r}\mathrm{d}\theta$ which is being convected due to particle self-propulsion  and which undergoes 
rotational diffusion and binary particle collisions, as given by the collision integrals $\mathcal{I}_\textit{d}[f]$ and $\mathcal{I}_\textit{c}[f,f]$, respectively; for explicit expressions please see the Supplemental Material~\cite{SUP}. Note here the critical difference to field theories for straight-moving particles~\cite{toner1995long,baskaran_hydrodynamics_2008,Mishra,BertinLong}; there is an additional angular derivative in the convection term, which reflects the fact that the particles are moving on circular orbits. In the following we rescale time, space and density such that $v_0{=}\lambda{=}d{=}1$. Then, the only remaining free parameters are the noise amplitude $\sigma$, $\kappa_0$, and the mean particle density   $\bar{\rho} {=} A^{-1} \int_{A} \mathrm{d}\vec{r} \int_{-\pi}^\pi\mathrm{d}\theta\,f(\vec{r},\theta,t)$ measured in units of $\lambda/(dv_0)$, \ie the number of particles found within the area traversed by a particle between successive self-diffusion events. 

To identify possible solutions of the Boltzmann equation and analyze their stability, we performed a spectral analysis. Upon expanding the one-particle distribution function in terms of Fourier modes of the angular variable, $\f_k(\vec{r},t){=}\int_{-\pi}^\pi\mathrm{d}\theta\,\e^{\imag \theta k}f(\vec{r},\theta,t)$, one obtains
\begin{align}
\label{Eq:BeqFT}
\partial_t \f_k
+&\frac{v_0}{2}\Bigl[\partial_x(\f_{k{+}1}{+}\f_{k{-}1}){-}\imag\partial_y (\f_{k{+}1}{-}\f_{k{-}1})\Bigr]{-}\imag kv_0\kappa_0\f_k\notag\\
&={-}\lambda(1{-}\e^{-(k\sigma)^2/2}) \f_k 
  {+}\summe{n}{-\infty}{\infty}\mathcal{I}_{n,k}\f_n \f_{k{-}n}\,,
\end{align}
where explicit expressions for the collision kernels $\mathcal{I}_{n,k}(\sigma)$ are given in the Supplemental Material~\cite{SUP}. For $k{=}0$, Eq.~(\ref{Eq:BeqFT}), yields the continuity equation $\partial_t \rho {=} {-}\vec{\nabla}{\cdot}\vec{j}$ for the local density $\rho (\vec{r},t) {:=} \f_0 (\vec{r},t)$ with the particle current given by $\vec{j} (\vec{r},t) {=} v_0(\Real\f_1,\Imag\f_1)^T$. In general, Eq.~(\ref{Eq:BeqFT}) constitutes an infinite hierarchy of equations coupling lower with higher order Fourier modes. 

\begin{figure}[t]
\flushleft
\centering
\includegraphics[width=\columnwidth]{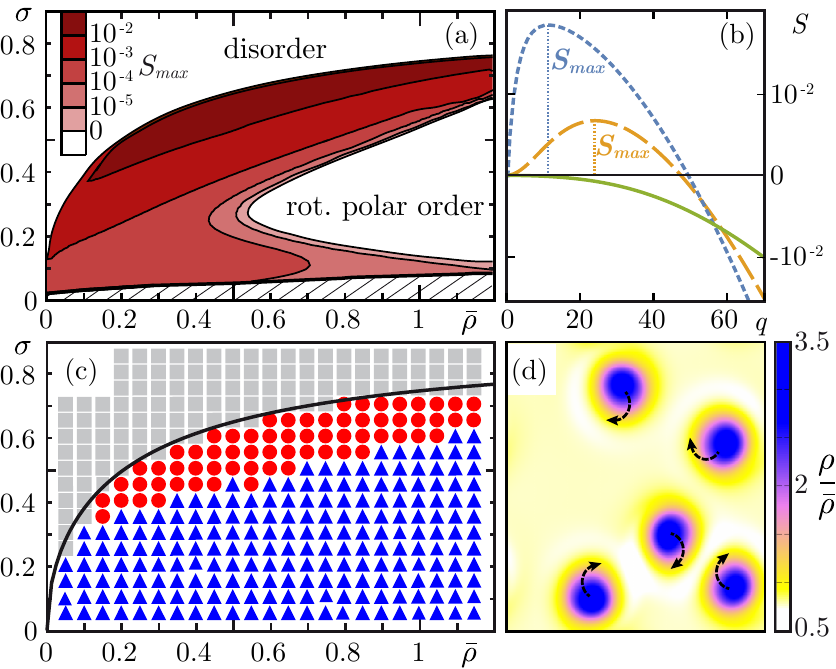}
\caption{\textbf{(a)} Stability of homogeneous solutions of Eq.~\eqref{Eq:BeqFT} as a function of $\sigma$ and ${\bar \rho}$ in units of $\lambda/(dv_0)$ . White and red areas denote regions where finite wavelength perturbations of the homogeneous solutions are stable and unstable, respectively. The color code denotes the value of the maximal growth rate $S_{max}$. \textbf{(b)} Dispersion relation of $S(q)$ ($q$ in units of $2\pi/\sqrt{A}$) for ${\bar \rho}{=}0.8$ and $\sigma{=}0.7$ (short dashed line), $\sigma{=}0.6$ (long dashed line) and $\sigma{=}0.4$ (solid line). Vertical lines indicate $S_\text{max}$. \textbf{(c)} Phase diagram for density ${\bar \rho}$ and $\sigma$ displaying phases of homogeneous disorder (gray rectangles), swirls (red circles) and homogeneous order (blue triangles). The solid line marks the analytic solution of $\rho_c(\sigma)$. An overlay of \textbf{(a)}, \textbf{(c)} can be found in the Supplemental Material~\cite{SUP}. \textbf{(d)} Snapshot of swirl patterns (${\bar \rho}{=}0.8$, $\sigma{=}0.7$). All swirls are moving clockwise on circular paths.}
\label{Fig::resultsBS}
\end{figure}

A linear stability analysis of Eq.~(\ref{Eq:BeqFT}) enables further progress. Since $\mathcal{I}_{n,0}{=}0$ for all $n$, a state with spatially homogeneous density ${\bar \rho} {=} f_0$ and all higher Fourier modes vanishing is a stationary solution to Eq.~\eqref{Eq:BeqFT} (\emph{disordered state}). To linear order, the dynamics of small perturbations $\delta\f_k$ with respect to this uniform state is given by $\partial_t \, \delta\f_k {=} \mu_k({\bar \rho},\sigma) \, \delta\f_k$, where $\mu_k ({\bar \rho},\sigma) {=} (\mathcal{I}_{0,k}{+}\mathcal{I}_{k,k}){\bar \rho} -\lambda(1{-}\e^{-(k\sigma)^2/2})$. For a polar collision rule, as considered here, only $\mu_1$ can become positive, defining a critical density $\rho_{c} (\sigma)$ at $\mu_1(\rho_{c},\sigma) {:=} 0$ [Fig~\ref{Fig::resultsBS}(a)]. Above threshold (${\bar \rho} {>} \rho_c$), the spatially homogeneous state is unstable, the particle current grows exponentially, and collective motion may emerge. 

In close proximity to the critical density $\rho_{c} (\sigma)$ a weakly non-linear analysis yields further insights into the dynamics of the system and the ensuing steady states. Here we follow Ref.~\cite{BertinBoltzmannGLApproach} and assume small currents $\f_1 {\ll} 1$ at onset. 
Then, balancing of the terms in the continuity equation, the equation for $\f_1$, and terms involving $\f_1$ in the equation for $\f_2$ implies the scaling ${\rho} {-} \bar{\rho} {\sim} \f_1$, $\f_2{\sim}\f_1^2$ as well as weak spatial and temporal variations $\partial_{x/y}{\sim}\f_1$, $\partial_t{\sim}\f_1$.
To include the lowest-order damping term in $\f_1$, we retain terms up to cubic order in $\f_1$. This 
yields the following hydrodynamic equation for the complex particle current $v_0 f_1 (\vec{r},t) {=} j_x (\vec{r},t) {+} \imag j_y(\vec{r},t)$
\begin{align}
\label{Eq:HydroEqu}
\partial_t \f_1 (\vec{r},t) 
=&\left[\alpha(\rho-\rho_c)
{+}\imag v_0\kappa_0\right]  \f_1 
{-} \xi|\f_1|^2\f_1 
{-} \frac{v_0}{2}\nabla \rho \notag \\ 
&{-}\beta\f^*_1 \nabla \f_1 
{-}\gamma\f_1\nabla^* \f_1
{+}\nu\nabla^* \nabla \f_1\,,
\end{align}
where $\nabla{:=}\partial_x{+}\imag\partial_y$. While this equation shows the same functional dependencies on local density and current as found in systems with straight propulsion~\citep{BertinLong}, the coefficients $\alpha$, $\xi$, $\nu$, $\gamma$ and $\beta$ are now \emph{complex-valued}
(for explicit expressions please see the Supplemental Material \cite{SUP}). This can be traced back to the angular convection term in Eq.~\eqref{Eq:Boltzmannactivematter}, or equivalently to the corresponding phase-shift term in Eq.~\eqref{Eq:BeqFT}. As a consequence, the field theory of active systems with particles moving on circular orbits with defined chirality is generically given by a \emph{complex Ginzburg-Landau (GL) equation with convective spatial coupling} as well as \emph{density-current coupling}. 
This constitutes a highly interesting generalization of the standard (diffusive) complex GL equations~\cite{AransonCGLGL,MoralesGinzLandau}, and is qualitatively different to real GL-type equations that were previously applied in the context of self-propelled particles~\cite{BertinBoltzmannGLApproach}. 
Above threshold, ${\bar \rho}{>}\rho_c(\sigma)$, the active chiral hydrodynamics described by the generalized GL equation Eq.~(\ref{Eq:HydroEqu}) exhibits a uniform oscillatory solution with $f_1 {=} F_1\e^{i\Omega_0t}$, \ie a state in which particles move on a circular (chiral) path with an angular velocity $\Omega_0{=}v_0\kappa_0{-}\alpha(\bar \rho{-}\rho_c)\text{Im}[\xi]/\text{Re}[\xi]$; the amplitude $F_1 {=} \left({\alpha(\bar \rho{-}\rho_c)}/{\text{Re}[\xi]}\right)^{1/2}$ gives the particle density.
However, a linear stability analysis of Eq.~(\ref{Eq:HydroEqu}) shows that for densities slightly larger than $\rho_c$ this oscillatory solution is linearly unstable against finite wavelength perturbations in the current and density fields. Preliminary numerical solutions of the generalized GL equation, Eq.~\eqref{Eq:HydroEqu}, take the form of rotating spots of high density that appear to show turbulent dynamics~\cite{SUP, upcoming}. This is qualitatively distinct from the high-density bands found for straight-moving particles~\cite{Chate2008,Chate2008a} and the vortex field of a fluid coupled to torque dipoles \cite{furthauer_active_2012,furthauer_active_2013}.

Far above threshold, closure relations such as those discussed above \cite{BertinBoltzmannGLApproach} may become invalid and with them the ensuing hydrodynamic equations. 
Therefore, we proceed with the full spectral analysis of the Boltzmann equation, Eq.~\eqref{Eq:BeqFT}, as detailed in the Supplemental Material~\cite{SUP}. First, we numerically calculate the spatially homogeneous solutions for all angular Fourier modes $\f_k$ below some cutoff wave vector $k_\text{max}$. For given values of ${\bar \rho}$ and $\sigma$ and a desired accuracy $\varepsilon$ of this mode truncation scheme, the cutoff is chosen such that $|\f_{k_\text{max}+1}| {<} \varepsilon$. We find that for ${\bar \rho}{<}\rho_c(\sigma)$ a spatially homogeneous state where all modes but $f_0$ vanish is the only stable state. In contrast, above threshold (${\bar \rho}{>}\rho_c(\sigma)$) there is a second solution for which $|\f_1|{>}0$. It corresponds to a polar ordered state whose orientation is changing periodically in time with frequency $v_0\kappa_0$. For moderate $\bar{\rho}{-}\rho_c$, the amplitude quantitatively agrees with the result from the generalized GL equation; see Supplemental Material~\cite{SUP}. 
In a second step, we consider wave-like perturbations, $\delta f_k (\vec{q})$ with wave-vector $\vec{q}$, of the spatially homogeneous oscillatory solution in a co-rotating frame. The largest real part of all eigenvalues of the corresponding linearized system for $\delta f_k$ then yields the linear growth rate $S(q)$ [Fig.~\ref{Fig::resultsBS}(b)]. In accordance with the linear stability analysis of Eq.\eqref{Eq:HydroEqu}, we find that for densities slightly larger than $\rho_c$ a spatially homogeneous solution is unstable against finite wavelength perturbations. The dispersion relation $S(q)$ exhibits a band of unstable modes, with the maximal growth rate $S_\text{max}$ decreasing as one moves away from the threshold $\rho_c$ [Fig.~\ref{Fig::resultsBS}(a,b)]. Actually, there is lobe-like regime in parameter space where $S(q){<}0$ [Fig.~\ref{Fig::resultsBS}(a)], and hence a homogeneously polar ordered state with rotating direction is stable. We  emphasize here that our stability portrait [Fig.~\ref{Fig::resultsBS}(a)] is independent of $\kappa_0$ and hence equally valid for systems of straight-moving particles. 
For our two approaches (Fig.~\ref{Fig::resultsBS}(a) and Fig.~\ref{Fig::resultsBD}(d)), the onset to order is governed by a similar trend~\cite{SUP}, common for active systems~\cite{Marchetti,RamaswamyReview2010}: disorder prevails for low density or high noise, order is promoted for high density or low noise.

To determine the spatiotemporal dynamics in the regime where neither a spatially homogeneous state nor a homogeneously polar ordered state are stable, we resort to a modified version of the SNAKE algorithm~\cite{FloSNAKE} to numerically solve Eq.~\eqref{Eq:Boltzmannactivematter}. It accurately reproduces the threshold value $\rho_c (\sigma)$ at which the spatially homogeneous state becomes unstable [Fig.~\ref{Fig::resultsBS}(c)]. 
Above threshold (${\bar \rho} {>} \rho_c$) we find that local density fluctuations quickly grow and evolve into stable swirls, \ie disc-like flocks of high density and polar order moving on circular paths; see Fig.~\ref{Fig::resultsBS}(d), and Movie 6 in Supplemental Material~\cite{SUP}. 
The radius of such a path is approximately given by $R_0$. 
These swirl patterns closely resemble the swirling flocks observed in the Brownian dynamics simulations for short polymer arc angles [Movie 5~\cite{SUP}], as well as our preliminary numerical solutions of the generalized GL equation, Eq.~\eqref{Eq:HydroEqu}~\cite{upcoming,SUP}. 
Moreover, in accordance with the spectral analysis, we find a second threshold density, above which the system settles into a homogeneously polar ordered state with a periodically changing orientation [Movie 7~\cite{SUP}]. 
The amplitude and frequency of the polar order agree with the numerical results of the spectral analysis to high accuracy~\cite{SUP}, while the numerically determined phase boundaries differ. The SNAKE algorithm produces stable swirl patterns only in a parameter regime where our linear stability analysis yields significant growth rates. This is mainly due to spurious noise caused by the discretization of the angular variable, which tends to suppress inhomogeneities in the regime of small growth rates. Furthermore, the finite system size constricts the band of possible modes and allows only for patterns of sufficiently short length scales.

For active systems of circling particles that interact via steric repulsion, our microscopic and mesoscopic treatments strongly suggests that a phase of collective vortex structures is a generic feature. Within this class, our work shows that extended polymers which as a whole follow circular tracks can form closed rings. Concerning our motivation of circling FtsZ, further research is needed to elucidate the dynamics of treadmilling; yet our minimal kinetic assumption suggests that varying particle density alone suffices to regulate the patterns as observed by Loose and Mitchison~\cite{loose_bacterial_2013}. 
Compared to systems of straight moving particles we find qualitatively new phenomena~\cite{SUP,upcoming}.
For those systems, it was already reported that (globally achiral) vortices can occur due to collisions of particles of asymmetric shape~\cite{wensink_controlling_2014} or due to memory in orientation~\cite{sumino_large-scale_2012,nagai2015}. 
Some of our findings, like the polymer length dependence of patterns and the possible emergence of active turbulence~\cite{dunkel_fluid_2013,bratanov_new_2015}, pose interesting questions for future work. Our analysis yields a mapping of the emergent dynamics onto a generalized Ginzburg-Landau equation, providing a connection between active matter and nonlinear oscillators~\cite{upcoming}.


\begin{acknowledgments}
We thank F. Th\"{u}roff, L. Reese, and J. Knebel for helpful discussions.
This research was supported by the German Excellence Initiative via the program `NanoSystems Initiative Munich' (NIM), and the Deutsche Forschungsgemeinschaft (DFG) via project B02 within the Collaborative Research Center (SFB 863) ``Forces in Biomolecular Systems''. 

J.D. and L.H. contributed equally to this work. 
\end{acknowledgments}


%

\cleardoublepage
\newpage

\pagebreak

\onecolumngrid
\begin{center}
\textbf{\large Supplemental Material: Active Curved Polymers form Vortex Patterns on Membranes}\\
\vspace{0.3cm}
Jonas Denk, Lorenz Huber, Emanuel Reithmann, and Erwin Frey\\
\textit{Arnold Sommerfeld Center for Theoretical Physics (ASC) and Center for NanoScience (CeNS), Department of Physics, Ludwig-Maximilians-Universit\"at M\"unchen, Theresienstrasse 37, 80333 M\"unchen, Germany}
\end{center}
\vspace{0.3cm}
\twocolumngrid

\setcounter{equation}{0}
\setcounter{figure}{0}
\setcounter{table}{0}
\setcounter{page}{1}
\makeatletter
\renewcommand{\theequation}{S\arabic{equation}}
\renewcommand{\thefigure}{S\arabic{figure}}
\renewcommand{\bibnumfmt}[1]{[S#1]}
\renewcommand{\citenumfont}[1]{S#1}

\section{Comment on Treadmilling}

In their experiments~\cite{Sloose_bacterial_2013}, Loose and Mitchison observe that FtsZ polymers undergo depolymerization and polymerization processes leading to an effective translation in the direction of the polymers' backbones. However, the underlying molecular details are unclear, as they involve many qualitatively and quantitatively unknown reactions and a yet unstudied interplay of different auxiliary proteins (\eg FtsA, ZipA). Here, we neglect these details and focus on the collective effects of many FtsZ polymers retaining only their effective movement along circular tracks. To realize this kind of motion we assume an intrinsic particle velocity.

\section{Numerical Implementation of Brownian Dynamics}
In the following, we discuss the details of the implementation of the Brownian dynamics simulations. We use a bead-spring model~\cite{Schirico_kinetics_1994,Swada_stretching_2007} that comprises the following discretization scheme: a polymer of length $L$ is subdivided into $N$ beads at positions $\vec{r}_i=(x_i,y_i)^T$ ($i=1,2,...,N$), with $N-1$ bonds of length $a$; the (normalized) bond vectors are given by $\partial_s \vec{r}\approx\frac{\vec{r}_{i+1}-\vec{r}_i}{a}=:\hat{\vec{t}}_i$; the bending angle between two adjacent bonds is given by $\theta_i=\arccos(\hat{ \vec{t}}_{i+1}\cdot\hat{\vec{t}}_i)$. The corresponding bending energy reads 
\begin{align}
E_{bend}=\frac{\ell_p}{2 a}k_BT\sum_{i=1}^{N-2}(\theta_i-\theta_0)^2.
\end{align}
where $\theta_0\approx a\kappa_0$ is the spontaneous bending angle.
In the bead-spring model, neighboring beads are  connected by stiff harmonic springs. The corresponding stretching energy is given by
\begin{align}
E_{stretch}=\frac{k}{2}\sum_{i=1}^{N-1}(|\vec{r}_{i+1}-\vec{r}_{i}|-a)^2.
\end{align}
In the simulations, the spring constant $k$ is chosen larger than all other force constants to account for the fact that biopolymers are nearly inextensible; as a consequence, stretching modes relax fast compared to other dynamic processes. At the same time, $k$ cannot be chosen arbitrarily large as this would strongly limit the maximal simulation time $T_{max}$ (see below for values).

In the two-dimensional system of $M$ polymers, we assume steric repulsion between adjacent polymer segments $\vec{r}_i^{(m)}$ ($m=1,2,...,M$). As an interaction potential we use a truncated Lennard-Jones potential~\cite{Sgoldstein_nonlinear_1995,Shinczewski_end-monomer_2009,Sbennemann_molecular-dynamics_1999}
\begin{align}
(E_{int})^{(mn)}_{ij}=\epsilon \left[\left(\frac{a}{r^{(mn)}_{ij}}\right)^{12}-\left(\frac{a}{r^{(mn)}_{ij}}\right)^6\right]\Theta(a-r^{(mn)}_{ij}),
\end{align}
with $r^{(mn)}_{ij}=|\vec{r}^{(m)}_i-\vec{r}^{(n)}_j|$, $\epsilon$ the potential strength, and $\Theta(r)$ the Heaviside step function. At distances smaller than the bond length $a$, the potential is strongly repulsive.

In the Langevin description, the equation of motion is given by a force balance between elastic, active, thermal and dissipative terms. For the $i$-th bead of a polymer, the equation of motion reads
\begin{align}\nonumber
\zeta \partial_t \vec{r}_i&=-\frac{\delta E}{\delta \vec{r}_i}+\vec{F}_i^{prop}+\boldsymbol{\eta}_i\\ \label{len}
&=\vec{F}_i^{bend}+\vec{F}_i^{stretch}+\vec{F}_i^{int}+\vec{F}_i^{prop}+\boldsymbol{\eta}_i
\end{align}
where $E=E_{bend}+E_{stretch}+E_{int}$, $\vec{F}_{prop}$ is the propulsive force and the amplitude of the thermal forces is given by $\langle\boldsymbol{\eta}_i(t)\cdot\boldsymbol{\eta}_j(t')\rangle=4k_BT\zeta\delta_{ij}\delta(t-t')$. The bending, stretching and interaction forces $\vec{F}_i^{bend},\vec{F}_i^{stretch},\vec{F}_i^{int}$ are obtained by variation of the corresponding energetic terms with respect to the position vector $\vec{r}_i$ \cite{Schirico_kinetics_1994,Swada_stretching_2007}. We employ the following implementation of the tangential propulsive force $\vec{F}^{prop}=\zeta v_0 \partial_s\vec{r}$:
\begin{align}
\vec{F}_i^{prop}=\zeta v_0\left\lbrace\begin{array}{cc} \hat{\vec{t}}_1 & i=1\\
(\hat{\vec{t}}_{i-1}+\hat{\vec{t}}_i)/2 & 1<i<N\\
\hat{\vec{t}}_{N-1} & i=N
\end{array}\right.
\end{align}

For the integration of Eq.~\eqref{len} we use an Euler-Maruyama iteration scheme~\cite{Skloeden2013numerical} with sufficiently small time steps $\Delta=0.0001\tau$ with the unit time $\tau=\zeta a^2/(k_BT)$. In our simulations, we used the following set of parameters:  $L=9a,\ell_p=100a,k=500k_BT/a^2,\epsilon=1k_BT,\theta_0=0.2,\zeta=1$ and a periodic system of area $A=60a\times60a$ (such that it can contain many consecutive polymer lengths). In the main text, the unit of length is set to $a=100\, nm$, such that $L=0.9\, \mu m,\ell_p= 10\, \mu m$ are roughly similar to FtsZ filaments. The noise strength ${\sigma}=k_BT\ell_p/(\zeta v_0 L^2)$ was varied as follows: we changed the temperature scale in the interval $k_BT\in [0,1]$ for $v_0=5$, and for $k_BT=1$ varied $v_0$ in the range $v_0\in [1,5]$. The maximal simulation times $T_{max}$ for all simulations in the main text were chosen such that the single polymer rotation time $\tau_R=2\pi/(\kappa_0v_0)$ is much smaller. We took $T_{max}>400\tau_R$ and $T_{max}>700\tau$ for our data to provide a sufficiently large sampling interval for both convective and diffusive motion. To consolidate the results, data were recorded for 10 independent simulation for each given set of parameters.

\section{Analysis of the pair correlation function}

\begin{figure}[htp]
\centering
\includegraphics[width=1.\columnwidth]{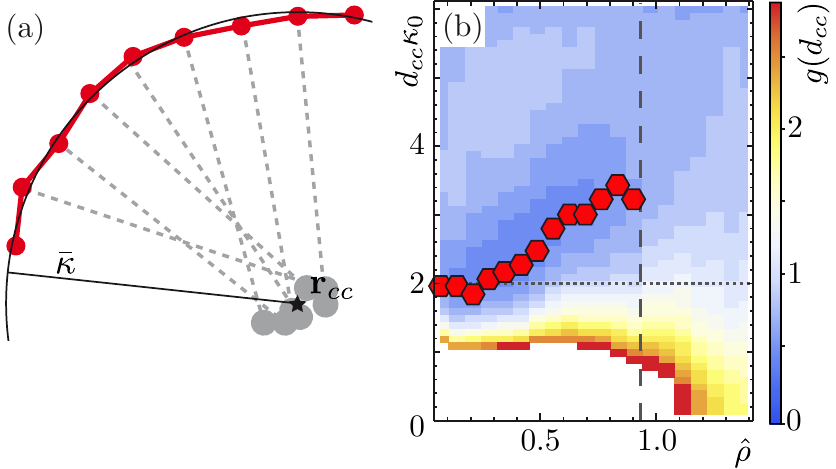}
\caption{\textbf{(a)} Illustration of the curvature center $\mathbf{r}_{cc}$ as determined by averaging over local centers with a mean contour curvature $\bar{\kappa}$ (polymer in red). \textbf{(b)} Heat map of the pair correlation function for ${\sigma}=0.247$ in terms of distances $d_{cc}$ and densities ${\rho}$. Red polygons denote the positions of $d_{cc}^{min}$. The short dashed line depicts the free polymer radius and the long dashed line marks the regime where $d_{cc}^{min}$ vanishes.}
\label{Fig::sfig1}
\end{figure}

To analyze the patterns observed in the Brownian dynamics simulations, we consider the pair correlation function $g(d_{cc})$~\cite{Sdhont_introduction_1996,Slu_molecular_2008} of center distances $d_{cc}=|\vec{r}^{(m)}_{cc}-\vec{r}^{(n)}_{cc}|$. The positions $\vec{r}^{(m)}_{cc}$ are the curvature centers of each polymer, generated by averaging over the local curvature and all local reference positions on a contour (see Fig.~\ref{Fig::sfig1}(a)). In contrast to the positions $\vec{r}^{(m)}$, the curvature centers do not oscillate due to self-propulsion and hence represent a more stable measure of particle position.

Figure~\ref{Fig::sfig1}(b) displays the contour of $g(d_{cc})$ for  parameters $k_BT=0.5$ and $v_0=5$ (\ie ${\sigma}=0.247$). For sufficiently small ${\rho}$, the density exhibits a local minimum at $d_{cc}^{min}$, the diameter of a vortex. This implies that there is a preferred vortex size and structure connected to the distance $d_{cc}^{min}$. These minima were determined after applying a Gaussian filter to suppress random fluctuation artifacts and then used to distinguish the observed patterns according to the 'phase' criteria introduced in the main text: \textit{disordered states} for $d_{cc}^{min}\approx2R_0$, \textit{vortex states} for $d_{cc}^{min}>2R_0$ and \textit{train states} without $d_{cc}^{min}$.

\section{Derivation of the hydrodynamic equations}

To assess the dynamics at larger scales, we employed a kinetic Boltzmann approach. The corresponding generalized Boltzmann equation for $f(\theta,\vec{r},t)$ is given by Eq.~\eqref{Eq:Boltzmannactivematter}. The self-diffusion and collision integrals $\mathcal{I}_\textit{d}$ and $\mathcal{I}_\textit{c}$, respectively, are given by
\begin{align}
\mathcal{I}_\textit{d}[f]&=\lambda\langle\int\limits_{-\pi}^{\pi}\mathrm{d}{\phi} f(\phi)\left[\delta(\theta-\phi-\eta)-\delta(\theta-\phi)\right]\rangle_\eta\, ,\\
\mathcal{I}_\textit{c}[f;f]&=\langle\int\limits_{-\pi}^{\pi}\mathrm{d}{\phi_1}\int\limits_{-\pi}^{\pi}\mathrm{d}{\phi_2} \mathcal{S}(|\phi_1-\phi_2|) f(\phi_1)f(\phi_2)\notag\\
&\times[\delta(\theta-\frac{1}{2}(\phi_1+\phi_2)-\eta)-\delta(\theta-\phi_1)]\rangle_\eta\,,
\end{align}
where $\mathcal{S}(\psi)=4 dv_0 |\sin(\frac{\psi}{2})|$ is the scattering cross section for spherical particles of diameter $d$ and velocity $v_0$ in two dimensions as detailed in Ref.~\cite{SBertinLong}. The collision integral represents ferromagnetic alignment of two particles with orientation $\phi_1$ and $\phi_2$ along their average angle $\theta=\frac{1}{2}(\phi_1+\phi_2)$. The brackets denote an average over a Gaussian-distributed noise variable $\eta$. To obtain a dimensionless form we used the rescaling
\begin{align*}
t &\rightarrow t\cdot \lambda^{-1}\notag\,,\\
\vec{x} &\rightarrow \vec{x}\cdot v_0\lambda^{-1}\notag\,,\\
f &\rightarrow f\cdot \rho_0 \,,\\
\kappa_0 & \rightarrow \kappa_0 \cdot v_0\lambda^{-1}\, ,
\end{align*}
with $\rho_0=\lambda/(dv_0)$. Measuring time, space and density in units of  $\lambda^{-1}$, $v_0\lambda^{-1}$, and $\rho_0$, respectively, allows to set $d=\lambda=v_0=1$. Then, the only remaining free parameters are the noise amplitude $\sigma$, $\kappa_0$, and the mean particle density $\bar{\rho} = A^{-1} \int_{A} \mathrm{d}\vec{r} \int_{-\pi}^\pi\mathrm{d}\theta\,f(\vec{r},\theta,t)$. To proceed, we performed a Fourier transformation of the angular variable: $\f_k(\vec{r},t)=\int_{-\pi}^\pi\mathrm{d}\theta\,\e^{\imag \theta k}f(\vec{r},\theta,t)$. This leads to the Boltzmann equation in Fourier space, Eq.~\eqref{Eq:BeqFT}, where the Fourier transforms $\mathcal{I}_{n,k}$ are given by
\begin{align}
\mathcal{I}_{n,k}=\int\limits_{-\pi}^{\pi}\!\frac{\mathrm{d}\Phi}{2\pi}\,\mathcal{S}(|\Phi|)\left[\Pft\cos(\Phi(n-k/2))-\cos(\Phi n)\right]\,.
\end{align}
$\Pft=\e^{-(k{\sigma})^2/2}$ is the Fourier transform (characteristic function) of the Gaussian noise with standard deviation ${\sigma}$. Note that $\mathcal{I}_{n,0}=0$ for all $n$. For $k=0$, Eq.~\eqref{Eq:BeqFT} hence yields the continuity equation $\partial_t \rho=-\frac{1}{2}(\nabla \f^*_{1}+\nabla^* \f_{1})=-\vec{\nabla}\cdot\vec{j}$ for the local density $\rho (\vec{r},t) := \f_0 (\vec{r},t)$ with the particle current given by $\vec{j} (\vec{r},t) =v_0(\Real\f_1,\Imag\f_1)^T$. In order to get a closed equation for the particle current at onset, we assume small currents $\f_1\ll1$ and use the truncation scheme: $\rho-\bar{\rho}\sim\f_1 $, $\partial_{x/y}\sim\f_1$, $\partial_t\sim\f_1$, $\f_2\sim\f_1^2$ with vanishing higher modes as presented for polar particles with ferromagnetic interaction in Ref.~\cite{SBertinBoltzmannGLApproach}. In analogy to Ref.~\cite{SBertinLong}, we retained only terms up to cubic order in $\f_1$ in the Boltzmann equation, Eq.~\eqref{Eq:BeqFT}, for $k=1$. The equation for $\f_1$ then couples to the nematic order field $\f_2$ via a term $\sim\f_1^*\f_2$ of order $\f_1^3$, where the star denotes complex conjugate. Writing down contributions from Eq.~\eqref{Eq:BeqFT} for $k=2$ of quadratic order in $\f_1$ yields an expression for $\f_2$ as a function of $\f_1$. The expression for $\f_2$ can then be substituted into Eq.~\eqref{Eq:BeqFT} for  $k=1$ to obtain a closed equation for $\f_1$. Together with the continuity equation, the hydrodynamic equations for the density and the particle current read
\begin{subequations}\label{HydroEquPolar}
\begin{align}
\partial_t \rho=&-\frac{1}{2}(\nabla \f^*_{1}+\nabla^* \f_{1})\,,\label{ContEqu}\\
\partial_t \f_1=&\left[\alpha(\rho-\rho_c)+\imag v_0\kappa_0\right] \f_1-\xi|\f_1|^2\f_1+\nu\nabla^* \nabla \f_{1}\notag\\
&-\gamma\f_1\nabla^* \f_1-\beta\f^*_1 \nabla \f_{1}-\frac{v_0}{2}\nabla \rho\,,\label{PolOrder}
\end{align}
\end{subequations}
where $\nabla:=\partial_x+\imag\partial_y$. The coefficients are given by 
\begin{flalign}
&\alpha:=(\mathcal{I}_{0,1}+\mathcal{I}_{1,1})\notag ,\\
&\rho_c=\frac{\lambda(1-\hat{P}_{1})}{\mathcal{I}_{0,1}+\mathcal{I}_{1,1}}\,\notag ,\\
&\nu:=-\frac{1}{4}\frac{1}{\lambda(\hat{P}_{2}-1)+2\imag v_0 \kappa_0+(\mathcal{I}_{0,2}+\mathcal{I}_{2,2})\rho}\,\notag ,\\
&\xi:=-4 (\mathcal{I}_{-1,1}+\mathcal{I}_{2,1}) \nu \mathcal{I}_{1,2}\,\notag ,\\
&\beta:=2 (\mathcal{I}_{-1,1}+\mathcal{I}_{2,1}) \nu\,\notag ,\\
&\gamma:=4\nu \mathcal{I}_{1,2}\,.
\label{coefficients}
\end{flalign}
We note that the employed truncation scheme implies fast relaxation of the nematic order field $\f_2$ such that $\partial_t\f_2$ is assumed to be negligible on time scales of the dynamics of $\f_1$.  $\f_2$ is then slaved to $\f_1$ via $\f_2=-2\nu\nabla \f_1+\gamma\f_1^2$.  

\subsection{Linear stability analysis}

For $\rho<\rho_c$ Eqs.~\eqref{HydroEquPolar} are solved by the homogeneous isotropic state:   $\rho=\rhohom=\textit{const.},\, \f_1=0$. For $\rho>\rho_c$ there is a second solution given by the homogeneous oscillatory state: $\rho=\rhohom,\,\f_1=F_1\e^{\imag \Omega_0 t}$ with $F_1=(\alpha(\rhohom-\rho_c)/\mathrm{Re}[\xi])^{1/2}$ and $\Omega_0=v_0\kappa_0-\alpha(\rhohom-\rho_c)\mathrm{Im}[\xi]/\mathrm{Re}[\xi]$. 

\subsubsection{Homogeneous isotropic state}

To study the stability of the homogeneous isotropic state we substitute $\rho=\rhohom+\delta\rho$ and $\f_1=\delta \f_1$ with the wave-like perturbations of the form
\begin{align}\label{Ansatzfouriermode}
\delta \rho(\vec{r},t)&\sim \delta \rho_\vec{q}\,\e^{\imag \vec{q}\cdot\vec{r}}\,\notag ,\\
\delta \f_1(\vec{r},t)&\sim \delta \f_{1,\vec{q}}\,\e^{\imag \vec{q}\cdot\vec{r}}\, ,
\end{align}
where $\delta\rho_\vec{q}$ and $\delta \f_{1,\vec{q}}$ are in general complex amplitudes that are assumed to be small. Periodic boundary conditions in our numeric solution impose $|\vec{q}|=n \frac{2\pi}{L},\,n\epsilon\mathbb{Z}$, where $L=\sqrt{A}$ and $A$ is the area of the (quadratic) system.
The linearized set of equations of motion for the perturbations $\delta\rho_\vec{q}(t)$, $\delta\f_{1,\vec{q}}(t)$ and $\delta\f_{1,\vec{q}}^*(t)$ has the characteristic polynomial
\begin{align}
\label{charpoly}
&- q^2 \alpha(\rhohom-\rho_c) + 	q^4 \Re[\nu]\notag\\
&+\left(2(\alpha(\rhohom-\rho_c)-\Re[\nu]q^2)^2+2(v_0\kappa_0-\Im[\nu]q^2)^2+q^2\right)S\notag\\
& + 4\left( -\alpha(\rhohom-\rho_c) +  \Re[\nu]q^2\right) S^2 + 2 S^3\,.
\end{align} 
where $S$ is the eigenvalue of the linearized set of equations for $\delta\rho_\vec{q}(t)$, $\delta\f_{1,\vec{q}}(t)$ and $\delta\f_{1,\vec{q}}^*(t)$. We note that $\Re[\nu]$ is positive for all densities. For $\rhohom<\rho_c$, all coefficients in~\eqref{charpoly}, including the $S$-independent terms are positive, such that~\eqref{charpoly} only yields $S$ with negative real part. Thus, for $\rhohom<\rho_c$ the homogeneous isotropic state is linearly stable against inhomogeneous wave-like perturbations. For $\rhohom-\rho_c>0$, the real part of $S$ becomes positive where the fastest growing mode is always at $q=0$. 

\subsubsection{Homogeneous oscillatory state}

\begin{figure}[ht]
\centering
\includegraphics[width=1.\columnwidth]{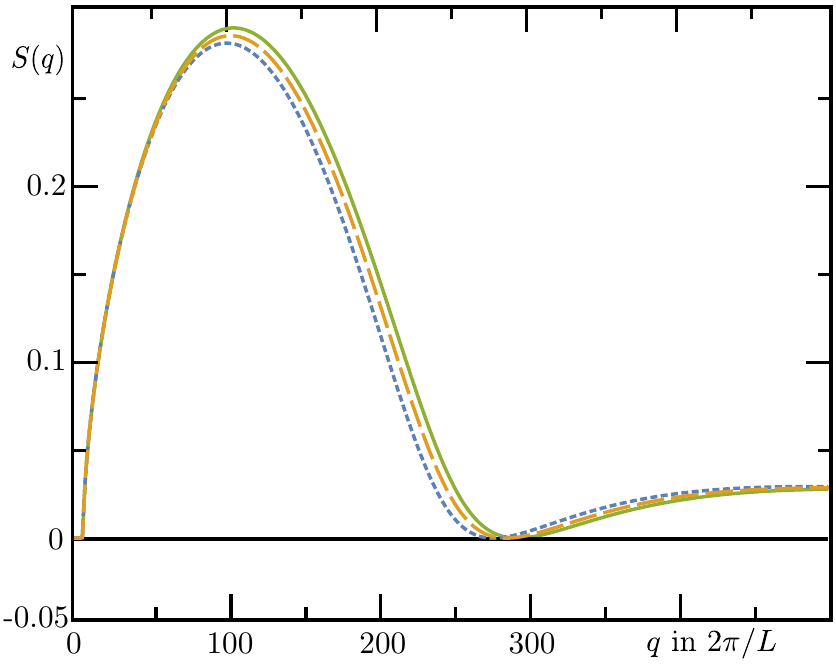}
\caption{Dispersion relations for $\sigma=0.6,\,0.4$ and $0.1$ (short-dashed, long-dashed and solid lines, respectively) at $\bar{\rho}=0.8$. }
\label{Fig::DispRelanalytic}
\end{figure}

To study the stability of the homogeneous oscillatory solution we substitute small perturbations in the basis of the homogeneous oscillating solution:
\begin{align}
\rho=&\rhohom+\delta\rho_{(0)}\notag\\
&+\sqrt{\frac{\alpha(\rhohom-\rho_c)}{\Re[\xi]}}\,\delta\rho_{(1)}\e^{\imag\Omega_0 t}+\sqrt{\frac{\alpha(\rhohom-\rho_c)}{\Re[\xi]}}\,\delta\rho_{(1)}^*\e^{-\imag\Omega_0 t}\notag, \\
\f_1=&F_1\e^{\imag\Omega_0 t}+\delta f_{(0)}\notag\\
&+\sqrt{\frac{\alpha(\rhohom-\rho_c)}{\Re[\xi]}}\,\delta f_{(1)}\e^{\imag\Omega_0 t}+\sqrt{\frac{\alpha(\rhohom-\rho_c)}{\Re[\xi]}}\,\delta f_{(2)}\e^{-\imag\Omega_0 t},
\end{align}
where the amplitudes $\delta\rho_{(0)},\,\delta\rho_{(1)},\,\delta f_{(0)},\,\delta f_{(1)}$ and $\delta f_{(2)}$ are again of the form \eqref{Ansatzfouriermode}. Truncating at the lowest order of $(\rhohom-\rho_c)$, which is $\sqrt{\alpha(\rhohom-\rho_c)}$, yields a closed set of linear equations for the amplitudes. The eigenvalue with the largest real part of this linear system determines the growth rate $S(q)$ of wave-like perturbations. We find that the dispersion relation yields positive $S(q)$ for finite $q$ (see Fig~\ref{Fig::DispRelanalytic}). 

\section{Numerical Linear Stability Analysis in the Full Phase Space}

In the derivation and the stability analysis of Eqs.~\eqref{HydroEquPolar} we rely on the assumption of small particle currents which might be justified at onset. However, this assumptions is in general questionable and not well justified for densities much larger than $\rho_c$. To obtain a stability map for the full phase space (Fig.~\ref{Fig::resultsBS}), we first calculated the homogeneous solution of Eq.~\eqref{Eq:BeqFT} retaining only modes up to $k_\text{max}$. Given some values of $\bar{\rho}$ and $\sigma$ and a desired accuracy $\epsilon$ of this mode truncation scheme the cutoff is chosen such that $|\f_{k_{max+1}}| < \epsilon$. As a next step, we linearized Eq.~\eqref{Eq:BeqFT} with respect to this solution and calculated the maximal growth rate $S(\vec{q})$ of wave-like perturbations with wave vector $\vec{q}$. If $S(\vec{q})>0$ for some $|\vec{q}|$, the homogeneous solution is unstable whereas if $S(\vec{q})<0$ for all $|\vec{q}|$, the corresponding homogeneous solution is stable.

\begin{figure}[ht]
\centering
\includegraphics[width=1\columnwidth]{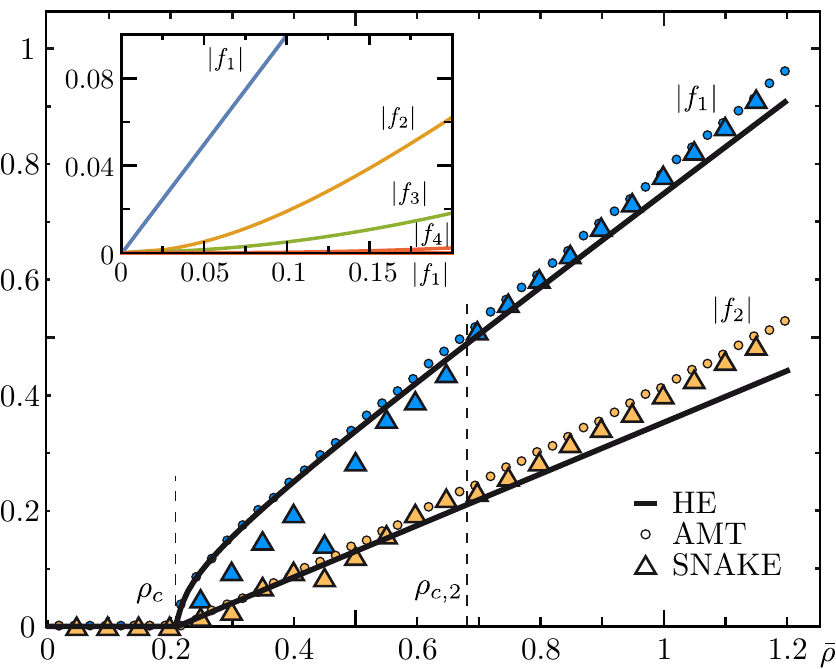}
\caption{Homogeneous solution for $\f_1$ and $\f_2$ for $\sigma=0.5$ obtained from the hydrodynamic equations Eqs.~\eqref{HydroEquPolar} (HE), the adapted mode truncation scheme (AMT), and the SNAKE algorithm. Note that within $\rho_c$ and $\rho_{c,2}$ (dashed vertical lines), the SNAKE algorithm yields swirl states and hence the corresponding mode values do not represent homogeneous states. The inset depicts the solutions for the first modes obtained from the AMT  and shows nonlinear scaling of higher modes with respect to $|\f_1|$.}
\label{Fig::Comparison}
\end{figure}

Note that the homogeneous version of Eq.~\eqref{Eq:BeqFT} (neglecting all gradient terms) is invariant under a phase shift $\f_k\rightarrow f_k\e^{\imag k v_0\kappa_0t}$. Choosing the orientation of the polar order at $t=0$ to be aligned along the $x$-axis, Eq.~\eqref{Eq:BeqFT} is solved by $\f_k=|f_k|\e^{\imag k v_0\kappa_0 t}$ with the time and space independent amplitude $|f_k|$. $|f_k|$ is then determined by the stationary homogeneous version of Eq.~\eqref{Eq:BeqFT}:
\begin{eqnarray}\label{Feq}
0=\lambda(\Pftnull-1) |f_k|+\summe{n}{-\infty}{\infty}\mathcal{I}_{n,k}\,|f_n| |f_{k-n}|\,.
\end{eqnarray}
This equation is identical to the stationary homogeneous Boltzmann equation for straight moving particles; i.e. where $\kappa_0=0$. Hence, the solutions for the amplitudes $|f_k|$ are identical to the solutions for the Fourier modes in systems of straight moving particles~\cite{SBertinLong}. To proceed, we truncate the infinite sum in Eq.~\eqref{Feq} at $k_\text{max}$ and calculate the solution of all $|f_k|$ with $|k|\leq k_\text{max}$. 
Fig.~\ref{Fig::Comparison} depicts the solution for the amplitude $|f_1|$ as compared to the solution of the generalized Ginzburg-Landau equation as well as the SNAKE algorithm. The explicit solution for $|\f_1|$ and higher modes justifies the scaling scheme used to derive Eqs.~\eqref{HydroEquPolar} in the vicinity of $\rho_c$ [Fig.~\ref{Fig::Comparison}, inset].
For decreasing noise $\sigma$ or increasing density $\bar{\rho}$ an increasing number of Fourier modes starts to grow [Fig.~\ref{Fig::Comparison}, inset]. In our numerical calculations we typically included $30-50$ Fourier modes. The dashed region in Fig.~\ref{Fig::resultsBS}(a) indicates the regime where we cannot find a nontrivial solution to Eq.~\eqref{Feq} by neglecting Fourier modes above the chosen $k_\text{max}=50$ and where we would have to choose a larger $k_\text{max}$.

With the substitution $\f_k=(|f_k|+\delta\f_k)\e^{\imag k v_0\kappa_0t}$ the linear system for $\delta\f_k$ then reads
\begin{eqnarray}\label{linstabnum}
\partial_t\delta\f_k=&-\frac{v_0}{2}(\nabla\delta\f_{k-1}+\nabla^*\delta\f_{k+1})+\lambda(\Pftnull-1)\delta\f_k\notag\\
&+\summe{n}{-\infty}{\infty}(\mathcal{I}_{n,k}+\mathcal{I}_{k-n,k})|f_{k-n}|\delta\f_n\,.
\end{eqnarray}
Here, we performed a coordinate transformation to a frame rotating with angular frequency $\kappa_0$ such that $\nabla\rightarrow\e^{\imag k v_0\kappa_0}\nabla$. 
Assuming wave-like perturbations as in Eq.~\eqref{Ansatzfouriermode}, we solved Eq.~\eqref{linstabnum} for the maximal eigenvalue and get the growth rate as a function of the wavenumber in the rotating frame (see Fig.~\ref{Fig::resultsBS}(b)). The maximum taken over all wavenumbers $|\vec{q}|>0$ then defines the maximal growth rate $S_{max}$ of wave-like perturbations. In agreement to previous results \cite{SBertinLong}, we found that the growth rate is maximal for $\vec{q}$ parallel to the particle current. The contour plot of $S_{max}$ as a function of $\bar{\rho}$ and $\sigma$ yields the phase diagram Fig.~\ref{Fig::resultsBS}(a). Note again, that our stability analysis and the resulting phase diagram Fig.~\ref{Fig::resultsBS}(a) is independent of curvature and also valid for the well-studied system of propelled particles without curvature~\cite{SChate2008a,SBertinLong,Sihle_kinetic_2011}. Hence, Fig.~\ref{Fig::resultsBS}(a) shows that the Boltzmann approach is capable of reproducing phases of all states observed in~\cite{SChate2008a,SChate2008} including a transition from travelling wave patterns to global homogeneous order. 

\section{Numerical Solution of the Boltzmann Equation with SNAKE}

\begin{figure}[ht]
\centering
\includegraphics[width=1.\columnwidth]{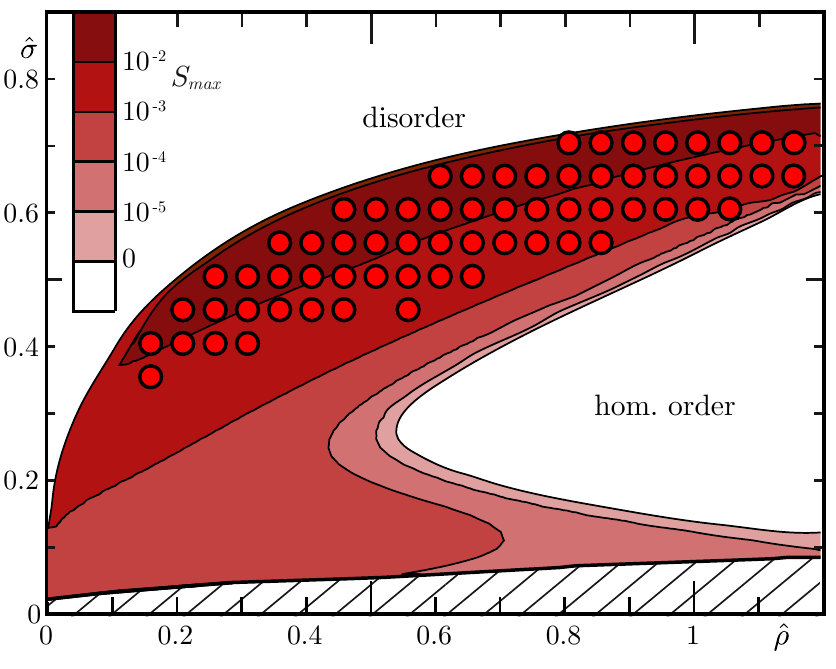}
\caption{Overlay of the parameter values where the SNAKE algorithm exhibits steady swirl patterns (red dots) together with the phase diagram obtained from the adapted mode truncation scheme (with $k_\text{max}=50$). In the shaded region, neglected Fourier modes become important.}
\label{Fig::GrowthrateContour}
\end{figure}

In order to study the resulting steady states in the regime where our linear stability analysis predicts inhomogeneities, we numerically solved the generalized Boltzmann equation, Eq.~\eqref{Eq:Boltzmannactivematter}. To this end we employed the SNAKE algorithm as introduced in Ref.~\cite{SFloSNAKE}. As tesselations we used a quadratic periodic regular lattice with equally sized angular slices. Circling propulsion was included by rotating the angular distribution of each lattice site with a frequency $v_0\kappa_0$ in addition to the straight convection steps. The system was initialized with a disordered state with small random density fluctuations around the mean density $\bar{\rho}=A^{-1}\int_{A}\rho(\vec{r},t)$. Changing $\kappa_0$ did not change the observed patterns qualitatively. In the limiting case of very small $\kappa$, we observed traveling wave patterns as reported in Refs.~\cite{SChate2008a,SChate2008,SFloSNAKE}. For Fig.~\ref{Fig::resultsBS}(c), Movie 6, and Movie 7 we used a lattice of of $200\times200$ grid points with lattice field size $2$ and angular disretization of $24$ angular slices; hence, $A=400\times400=160000$. In the swirl phase the swirl size grows for growing $\bar{\rho}-\rho_c$ whereas the radius of a swirl's motion stays at approximately $\kappa_0^{-1}$. Fig.~\ref{Fig::GrowthrateContour} shows the parameter values of $\bar{\rho}$ and $\sigma$ where the SNAKE algorithm exhibits steady swirl patterns together with the phase diagram obtained from the adapted mode truncation scheme.

\section{Remark on the shape of the phase curves}

When comparing the transition to order in the phase diagrams~\ref{Fig::resultsBD} and~\ref{Fig::GrowthrateContour} it should be noted that our particle-based and continuum approaches are distinct in the following features: polymer fluctuations vs. effective diffusion, multi-particle collisions vs. binary alignment, extended polymers vs. point particles.
The functional form of $\rho_c(\sigma)$~\eqref{coefficients} depends on the choice of diffusion and collision noise (e.g. equally Gaussian distributed). In contrast, the form of the transition line in our Brownian dynamics simulations depends on the choice of the phenomenological criteria (\textit{disordered states} for $d_{cc}^{min}\approx2R_0$, \textit{vortex states} for $d_{cc}^{min}>2R_0$ and \textit{train states} without $d_{cc}^{min}$). These differences result in different shapes of the phase boundaries. In addition, the observed patterns in the vortex phase are distinct. While for our particle-based model we find closed, rotating rings, dense, rotating swirls are observed in the continuum model (Fig.~\ref{Fig::resultsBD}(b) and Fig.~\ref{Fig::resultsBS}(d)). These differences are interesting and should be considered as part of the results we obtained. For example, these differences will guide future model building for specific models, e.g. the dynamics of FtsZ, as they emphasise what molecular details need to be accounted for. For the discussion of this work, however, our emphasis was on the topology of the phase diagram (similar trend of the onset to order) and the fact that in both models one finds a vortex phase.

\section{Movie descriptions}
\textbf{Movie1.mp4}: Brownian dynamics simulation of a system with $M=10$ polymers with $v_0=5, k_BT=1$ and hence ${\rho}=0.069,\, {\sigma}=0.247$.

\textbf{Movie2.mp4}: Brownian dynamics simulation of a system with $M=80$ polymers with $v_0=5, k_BT=1$ and hence ${\rho}=0.556,\, {\sigma}=0.247$.

\textbf{Movie3.mp4}: Brownian dynamics simulation of a system with $M=200$ polymers with $v_0=5, k_BT=1$ and hence ${\rho}=1.389,\, {\sigma}=0.247$.

\textbf{Movie4.mp4}: Brownian dynamics simulation with parameters as in Movie 3, except for a changed curvature angle $\theta_0=0.333$, resulting in an polymer arc angle $L\kappa_0=3$. 

\textbf{Movie5.mp4}:  Brownian dynamics simulation with parameters as in Movie 3, except for a changed contour length $L=6$, resulting in an polymer arc angle $L\kappa_0=1.2$.  

\textbf{Movie6.mp4}: SNAKE solution for $\bar{\rho}=0.2$ and ${\sigma}=0.45$ with $\kappa_0=0.1$. The colour code denotes the local density $\rho/\bar{\rho}$. The orientation and length of the arrows indicates the orientation and amplitude of the local particle current. 

\textbf{Movie7.mp4}: SNAKE solution for $\bar{\rho}=0.75$ and ${\sigma}=0.2$ with $\kappa_0=0.1$. The colour code denotes the local density $\rho/\bar{\rho}$. The orientation and length of the arrows indicates the orientation and amplitude of the local particle current.

\textbf{Hydroswirl.mp4}: Preliminary results of the explicit integration~\cite{Supcoming} of the hydrodynamic Eqs.~\eqref{HydroEquPolar}. The video shows the time evolution of the density field $\rho(\vec{r},t)$, for parameters close above threshold $\bar\rho>\rho_c$. The system size is $A=80\times80=640$, $\bar\rho=0.5$, $\sigma=0.6$, and $R_0=5$.


\end{document}